
\magnification=\magstep1
\baselineskip=16pt plus 2pt

\parindent=12pt

\voffset=1.0truecm
\hsize=6truein
\vsize=8.5truein

\hoffset=0.5truecm

\line{\hfill TUM--HEP--230/95}
\line{\hfill SFB--375/25}
\line{\hfill hep--ph/95mmxxx}
\line{\hfill November 1995}

\bigskip\bigskip
\bigskip

\centerline{\bf PHENOMENOLOGICAL ASPECTS OF SUPERSYMMETRY}
\bigskip\bigskip
\bigskip
\centerline{\bf Hans Peter Nilles\footnote*{Lecture given at the
conference "Gauge Theories, Applied Supersymmetry and Quantum Gravity",
Leuven, Belgium, July 1995.}}

\medskip\bigskip
\centerline{\it Physik Department}
\centerline{\it Technische Universit\"at M\"unchen}
\centerline{\it D-85747 Garching, Germany}
\medskip
\centerline{\sl and}
\medskip
\centerline{\it Max-Planck-Institut f\"ur Physik}
\centerline{\it --Werner-Heisenberg-Institut--}
\centerline{\it D-80805 M\"unchen, Germany}
\bigskip\bigskip\bigskip
\centerline{ABSTRACT}
\medskip
{\begingroup

{\parindent=0pt
We discuss the possible applications supersymmetric theories might find
in the field of elementary particle physics. The supersymmetric
generalization of the $SU(3)\times SU(2)\times U(1)$ standard model is
discussed in detail. Special attention has been devoted to the
question of gauge coupling constant unification in the framework of
supersymmetric grand unified models.}

\endgroup}

\vfill\eject

\line{\bf 1. Introduction\hfill}
\medskip

The beautiful structure of supersymmetric theories is certainly one
of the reasons they attracted and still attract so much attention.
It remains to be seen, however, whether these theories have direct
applications in the field of particle physics.
In these lectures we shall discuss possible  supersymmetric
extensions of the standard model of particle physics [1] and the reasons
we think why such a generalization should be considered. We shall see
that phenomenological  considerations restrict the possible
realization of supersymmetry substantially. At energies below 100 GeV,
we know that supersymmetry is badly broken, but still at some higher
energies the world might be supersymmetric.
Before embarking on this trip to the supersymmetric world let us first
review shortly the basics of the nonsupersymmetric standard model.

   The standard model is based on the gauge interactions of the
strong and electroweak forces with gauge group
$SU(3)\times SU(2)\times U(1)$. It
thus contains 12 spin 1 gauge bosons: 8 gluons of $SU(3)$, 3 $SU(2)$ weak
gauge bosons and the hypercharge gauge boson of $U(1)$. The photon will
be a particular combination of the neutral $SU(2)$ gauge boson and the
hypercharge boson. The fermions of the theory consist of three
generations of quarks and leptons, where we assume the existence of
the top quark for which direct experimental evidence ist still
lacking. The spin-1/2 fermions of a family have the following
transformation properties with respect to
$SU(3)\times SU(2)\times U(1)$:
$$\eqalign{U^a={u\choose d} &= (3,2,1/6)\cr
                     \bar u &= (\bar 3,1,-2/3)\cr
                     \bar d &= (\bar 3,1,1/3)\cr
       L^a={\nu_e\choose e} &= (1,2,-1/2)\cr
                     \bar e &= (1,1,1)\cr}\eqno(1.1)$$
where $a = 1,2$ is an $SU(2)$ index and the first two entries in the
brackets denote the dimensions of the $SU(3)\times SU(2)$ representations
while the last entry denotes $U(1)$ hypercharge. Electric charge is
given by $Q=T_3  + Y$. Thus the up-quark, for example, has
$Q(u) =1/2 +1/6=2/3$
whereas for the down quark we obtain $Q(d) = - 1/3$.

   The so-called Higgs sector contains a scalar $SU(2)$-doublet
$$h={h^0\choose h^-} = (1,2,-1/2)\eqno(1.2)$$
with potential $V=\mu^{2}(h^{\dagger} h)+\lambda (h^{\dagger} h)^2$
and one also introduces Yukawa couplings for the interactions of the
scalars with the fermions
$$L_Y =g_d Uh\bar d + g_e Lh\bar e + g_u Uh^{\dagger}\bar u\eqno(1.3)$$
in all combinations that are allowed by
$SU(3)\times SU(2)\times U(1)$ gauge
symmetry. A spontaneous breakdown of $SU(2)\times U(1)$ occurs for
negative $\mu^2$
and the neutral component of  $h$  receives a vacuum expectation value (vev)
$$<h>={1\over{\sqrt 2}}{v\choose 0}\eqno(1.4)$$
where $v=(-\mu^2/\lambda)^{1/2}$. $SU(2)\times U(1)_Y$ is
broken to $U(1)_Q$   and three gauge bosons become massive
$$\eqalign{M_{W^\pm} &={1\over 2}g_2 v\cr
           M_Z       &={1\over 2} v \sqrt{g_1^2 +g_2^2}}\eqno(1.5)$$
where $g_1$ and $g_2$ are the coupling constants of $SU(2)$ and $U(1)$,
respectively. The $U(1)$ gauge coupling constant is given by
$$e=g_2\sin\theta_W=g_1\cos\theta_W\eqno(1.6)$$
where $\theta_W$  denotes the weak mixing angle. The mass of the physical
Higgs-scalar is given by $\sqrt{-2\mu^2}$.
   Yukawa couplings then allow, in presence of the spontaneous
breakdown of $SU(2)\times U(1)$, mass terms for the fermions.
The term $g_d hU\bar d$,
e.g. leads to $g_d vd\bar d=m_d d\bar d$.
 The masses and mixings for the three
families of quarks and leptons are parametrized by the $3\times 3$
Kobayashi-Maskawa[2] matrix.

   Let us now count the parameters of the model. We have three
gauge couplings $g_1$, $g_2$ and $g_3$    usually parametrized by
$\alpha_{\rm e.m.}$, $\alpha_{\rm strong}$
and $\sin\theta_W$. In the gauge sector we have in addition a
$\Theta$ -parameter
multiplying a $F^{\mu\nu}F^{\rho\sigma}\epsilon_{\mu\nu\rho\sigma}$
   in the action. Its actual value seems to
by very close to zero as can be deduced from the absence of the
electric dipole moment of the neutron. Nonetheless we have to treat
$\Theta$ as an arbitrary parameter and it still has to be understood why its
value is so small.

   In the Higgs sector we have introduced two parameters $\mu^2$
and $\lambda$ of which one
combination  defines the scale of $SU(2)\times U(1)$ breakdown
while the other determines the Higgs mass. The 9 fermion masses (not
including the possibility for neutrino-Majorana masses) are
parametrized by the Yukawa couplings. The same applies to quark
mixing consisting of 3 angles and one phase in the Kobayashi-Maskawa
matrix, the latter giving rise to CP-violation. We do not  know yet
whether there is a corresponding mixing in the lepton sector. In any
case we can conclude that the above mentioned quantities are
completely free parameters in the standard model. Any attempt to
understand their specific values will require a generalization of the
model. Apart from these questions we have eventually also to address
the more fundamental puzzles out of which I shall mention some in the
following. Why is the gauge group
$SU(3)\times SU(2)\times U(1)$, why is $SU(2)$
broken and why at a scale of 100 GeV and not at the Planck mass? Why
is the mass of the proton 1 GeV and is this scale related to other
physical scales? Why do we have this repetition of families, why 3
families and why does a family not contain exotic representations of
$SU(3)\times SU(2)\times U(1)$ (like e.g. a 3 of $SU(2)$)?
Why are neutrinos massless
(are they?) and why is the electron mass so small compared to the
$W$-mass? These and many more related questions are the subject of
discussions of the physics beyond the standard model.

   One important property of the standard model is the chirality of
the fermion spectrum. Fermion masses are protected by
$SU(2)\times U(1)$, i.e.
they can be nonzero only after $SU(2)\times U(1)$ breakdown. Thus all
fermion masses are proportional to the vev of the
Higgs-field (1.4) and this explains
why fermion masses cannot be very large compared to $M_W$. It does, of
course, not explain why the mass of the electron is so small compared
to $M_W$ and also the smallness of neutrino masses remains a mystery.
Only the top quark seems to be as heavy as allowed by
$SU(2)\times U(1)$. We
will regard this chirality of fermions as a very important property
of the standard model and will therefore in the course of these
lectures only discuss extensions that share these remarkable properties.

   Another important symmetry of the standard model is baryon (B)-
and lepton (L)- number conservation. From the requirement of gauge
invariance and renormalizability (i.e. absence of nonrenormalizable
terms in the action) the model has automatic B and L conservation.
Among other things this implies the stability of the proton. Possible
violations could come from higher dimensional (nonrenormalizable)
terms as e.g. four-fermion operators.
These operators have dimension 6 and therefore the coefficient $1/M_x^2$
has the dimension of inverse (mass)$^2$. $M_x$  denotes the scale of the new
physics that is responsible for proton decay. From the long lifetime
of the proton we conclude that $M_x$ must be larger than $10^{15}$ GeV, a very
large scale. For other processes, like lepton number violation, the
corresponding scale could still be in the TeV region. It is a central
question in all discussions of the physics beyond the standard model
to isolate these new processes and discuss the corresponding scales.

\bigskip\bigskip

\line{\bf 2. Why supersymmetry\hfill}
\medskip

The standard model contains a dimensionful scale of the order of 100 GeV,
represented by the masses of the intermediate gauge bosons. All parameters
of dimension mass in the model are related to the vev of the scalar field
that is responsible for the breakdown of $SU(2)\times U(1)$. If this would
be the only scale in physics we could regard this scale then as
{\it the} input parameter in the model and derive all mass parameters from it.
There are reasons to believe, however, that there exist other fundamental
scales in physics such as the Planck scale around $10^{19}$ GeV related to
the gravitational interactions or a hypothetical grand unified scale of
$10^{16}$ GeV in connection with the possible unification of strong and
electroweak interactions. Compared to these scales the weak scale is tiny,
in fact so tiny that one would think that one should find an explanation
for this fact. Such a reason could be a symmetry as we encountered in the
discussion of fermion masses, where chiral symmetry protected the masses.
Chiral symmetry cannot forbid scalar masses and can therefore not explain
the smallness of the weak scale.

Let us discuss this situation in detail. Recall the Higgs potential
$$V(h)=\mu^2\vert h\vert^2+\lambda\vert h\vert^4.\eqno(2.1)$$
The Higgs mass is $m =\sqrt{-2\mu^2}$ and
$M_W=g_2 <h>\approx 80$ GeV.
Experimental bounds on $m$ come from LEP
$m\geq 60$GeV while an upper bound
of 1 TeV can be argued from unitarity constraints. Observe that the
mass scale of the standard model $M_W$  is solely set by the
parameters $\mu^2$  and $\lambda$
in the Higgs sector.

  Theoretically the model is very appealing; it is not just based on
an effective Lagrangian, like e.g. the Fermi theory of weak
interactions, but it is a renormalizable field theory. This has
drastic consequences for the possible range of validity of the model;
would it be nonrenormalizable it necessarily would only be defined
with a cutoff $\Lambda$ (of dimension of a mass) and its region of validity
would be bounded from above by $\Lambda$. Above
$\Lambda$ one expects new things to
happen which are not described by the model. Since the standard model
is renormalizable it could, however, be valid in a much larger energy
range. Strangely enough this very nice property of the model
constitutes one of its problems. The mass scale of 100 GeV is put in
by hand and there is no understanding of its origin: it is a completely free
input parameter. In a more complete theory one would like to
understand the origin of $M_W$ in terms  of more fundamental
parameters like e.g. the Planck scale $M_P\sim 10^{19}$ GeV, but such a
complete theory would need more structure than present in the
standard model.

   A reconfirmation of the statement that $M_W$  is a completely free
parameter is found in the discussion of perturbation theory. The
parameter $\mu^2$ in (2.1) receives a contribution at the one loop level
which is quadratically divergent. There is nothing wrong with
quadratic divergencies as they do not spoil the consistency of the
theory; we regularize them and define the theory in terms of the
renormalized parameters. The actual correction to $\mu^2$  depends on the
regularization scheme and the renormalized quantity is an arbitrary
parameter even if we would have understood its value at the tree
level. This is true for all quadratically divergent quantities. These
divergences introduce a new mass scale in the theory which has nothing
to do with the scales already present;  it is an
arbitrary parameter which we can choose at our will. To understand
the origin of these masses the quadratic divergencies have to be
absent i.e. they have to be cut off at a larger scale by a new
physical structure. With such a {\it physical} cutoff
$\Lambda$ we would have
$$\delta\mu^2\sim\lambda\Lambda^2\eqno(2.2)
$$
and to understand the order of magnitude of $\mu^2$ it would not
be appropriate to have $\Lambda$ of the order of the Planck mass $M_P$ but
rather in the TeV region. An understanding of the order of magnitude
of $M_W$  would therefore require new physics in the TeV-region.

   Having agreed that the standard model might have this subtle
theoretical problem one has to look for ways out. The presence of
quadratic divergencies is originated by the existence of fundamental
scalar particles. One way out is to remove these scalars from the
theory. Since we have to break $SU(2)\times U(1)$ spontaneously (and want to
maintain Lorentz invariance) some scalar objects have to exist; they
could be composite as postulated in the technicolour approach[3]. A new gauge
interaction becomes strong in the region of a few hundred GeV;
leading to the formation of
condensates and many composite bound states. This is the new physics
in the TeV-region.

   But this is not the only possible solution and we could try to
insist to live with
fundamental scalar particles. Remember for this purpose the situation
with spin 1 particles. Models containing spin 1 particles have
usually serious theoretical problems unless there is a gauge symmetry
that makes these fundamental spin 1 particles acceptable. Observe that
this gauge symmetry also stabilizes the mass of these spin 1 particles; in
the symmetric limit they have to vanish. Could we also have such a
situation for scalar masses? In the standard model, of course, such a
situation is not present. We can take the limit
$\mu^2\rightarrow 0$      and this does
not enhance the symmetry of the action.

   The only known way to protect scalar masses is supersymmetry. This
symmetry relates bosons and fermions and therefore makes bosons as
well behaved as fermions, which implies the absence of quadratic
divergencies. Supersymmetry provides us with the physical cutoff
discussed earlier. In addition to the contribution to $\mu^2$ through
a scalar loop we have now a contribution of  the
supersymmetric partner of the Higgs boson propagating in the loop. In the
supersymmetric limit these
two contributions cancel exactly. If supersymmetry is broken the
masses of the boson-fermion multiplet are split. We get a contribution
$$\delta\mu^2\approx \lambda(m_B^2-m_F^2)\eqno(2.3)$$
and we would require the quantity on the right-hand side to be in
the TeV range. If we would remove the partner with mass $m_F$   from the
theory we would again recover the quadratic divergence of the
standard model. Thus to solve the Higgs problem we have to consider
new (supersymmetric) structure in the TeV-region.

\bigskip\bigskip

\line{\bf 3. The particle content of the supersymmetric standard model\hfill}
\medskip

Let us now start the construction of the supersymmetric generalization of the
standard model. I shall assume that the reader is familiar with the
concept of global supersymmetry as provided e.g. in [4]
or a previous review [5].

   We recall the particle content of the standard model.
Apart from the gauge bosons $G_\mu^a$, $W_\mu^i$, $B_\mu$ in the adjoint
representation we have quarks and leptons in three families with
quantum numbers
$$\eqalign{Q&={u\choose d}=(3,2,1/6)\cr
      \bar u&=(\bar 3,1,-2/3)\cr
      \bar d&=(\bar 3,1,1/3)\cr
           L&={\nu_e \choose e}=(1,2,-1/2)\cr
      \bar e&=(1,1,1)\cr}\eqno(3.1)$$
together with a Higgs doublet
$$h={h^0\choose h^-}=(1,2,-1/2)\eqno(3.2)$$
The spectrum of this model is not supersymmetric and we have to add
new degrees of freedom. There are no fermions in the adjoint
representation of $SU(3)\times SU(2)\times U(1)$ and we thus have to add gauge
fermions (gauginos), which together with the gauge bosons form a
massless vector superfield $V=(V_\mu,\lambda,D)$. Quarks and leptons require
spin 0 partners in chiral superfields e.g.
$\bar E = (\varphi_{\bar e},\bar e,F_{\bar e})$ where $\varphi_{\bar e}$
  is a complex scalar with  $\bar e$  quantum numbers. Next observe that the
lepton doublet has the same quantum numbers as the Higgs: could it
be that $\varphi_e=h^-$? Unfortunately it does not work. One reason is the
absence of lepton number violation and other reasons
will become clear in a moment. We thus have to add scalar
partners to all quarks and leptons. To the Higgs scalar we have to
join the partner spin 1/2 fermions. With these fermions $SU(2)\times U(1)$
is no longer anomaly free and we have to add a second Higgs chiral
superfield $\bar H=(1,2,+1/2)$. In short, every particle in the standard
model requires a new supersymmetric partner and one has to add a
second Higgs superfield.

   To construct the Lagrangian we first write the kinetic terms and
the gauge couplings in the usual supersymmetric way.
We still have to discuss the superpotential which
contains mass terms and the supersymmetric generalization of the
Yukawa couplings. If we write the most general superpotential
consistent with the symmetries and renormalizability it will contain
two sets of terms
$$g=g_{\rm w}+g_{\rm u}.\eqno(3.3)$$
Let me first discuss the term
$$g_{\rm w}=\mu H\bar H+g_E^{ij}L_i^aH^b\epsilon_{ab}\bar E_j+
g_D^{ij}Q_i^aH^b\epsilon_{ab}\bar D_j+
g_U^{ij}Q_i^a\bar H_a\bar U_j\eqno(3.4)$$
where $i, j = 1,\ldots 3$ is a family index and $a,b$ are $SU(2)$ indices
(colour indices are suppressed). It is not really clear whether we
want $\mu$  from a theoretical point of view but we need it to break
certain global symmetries that might be problematic. I will come back
to this point later. Observe  that we really need two Higgs
superfields to give masses to all quarks and leptons. We can here no
longer couple the up-type quarks to $h^*$ as we did in the nonsupersymmetric
case. It is then also clear that in the breakdown of $SU(2)\times U(1)$
both Higgses have to aquire a vev to provide masses to all quarks and
leptons.

  Unlike in the standard model where the requirement of gauge
symmetry and renormalizability automatically led to baryon and lepton
number conservation we are here not in such a nice situation. This
comes from the fact that the Higgs and the
lepton doublet superfields have the same
$SU(3)\times SU2)\times U(1)$ quantum numbers. Consequently we have additional
terms in (3.3) that we can write as (forgetting family indices)
$$g_{\rm u}=Q^aL^b\epsilon_{ab}\bar D+L^a\bar E L^b\epsilon_{ab}+
\bar U\bar D\bar D.\eqno(3.5)$$
These terms violate baryon and lepton number explicitly and lead to
proton decay mediated by the exchange of the scalar partner of the
d-quark. The rate for this process is unacceptaably large
as long as we assume the partner
of the d-quark to be lighter than the grand unification scale. Thus some of
the terms in (3.5) have to be forbidden.  Let us try  to achieve this with help
of a symmetry. We can turn the question the other way around. Suppose
we drop (3.5) from the superpotential; does the symmetry increase? In fact
it does. The new symmetry is a global symmetry that, however, does
not commute with supersymmetry (called $R$-symmetry[6]). Different
components in the same supermultiplet have different charges. The
concept of $R$-symmetry can best be explained in superspace. Suppose we
have a symmetry that transforms $\theta$ to
$e^{i\alpha}\theta$; so $\theta$  has charge $R=1$.
Suppose we have a chiral superfield $\phi$ transforming also with $R=1$.
 Then it is obvious that the scalar component transforms as
$$\varphi\rightarrow e^{i\alpha}\varphi\eqno(3.6)$$
with $R = 1$. But what happens to the fermion?
Since $R(\phi) = 1$ we have
$$(\theta\psi)\rightarrow e^{i\alpha}(\theta\psi)\eqno(3.7)$$
but the phase comes already from the $\theta$ transformation and obviously
$R(\psi) = 0$. The  $F$-component of the superfield has $R(F) = -1$.
Invariance of the Lagrangian requires
$\int{\rm d^2}\theta g$ to have $R = 0$ whereas ${\rm d^2}\theta$
  transforms with $R = -2$. In the given example only the term $\phi^2$ is
allowed in the superpotential. So far our discussion of the
implication of R-symmetry on chiral superfields. The vector
superfield is real and consequently $R = 0$. From this we conclude
$$\eqalign{R(V_\mu)&=0\cr
R(\lambda)&=1\cr}\eqno(3.8)$$
and this is a general and important statement. Gauginos transform
nontrivially under any $R$-symmetry. The R-symmetry, in particular,
forbids Majorana masses for the gauge fermions.

 Let us now go back to the superpotential (3.4) and (3.5). There is
an $R$-symmetry with e.g. $R(\theta)=1$ and
$$\eqalign{R(H,\bar H)&=1\cr
R(Q,L,\bar U,\bar D,\bar E)&=1/2\cr}\eqno(3.9)$$
which leaves $g_{\rm w}$ in (3.4) as the most general superpotential.
In other words this means that if we drop the terms in (3.5) a
continuous global $R$-symmetry appears. To forbid these terms in
principle a smaller symmetry like $R$-parity
$$R_p=(-1)^{3B+L+2S}\eqno(3.10)$$
(where $B$, $L$ are baryon, lepton number and  $S$  is the spin) would be
sufficient, but here a continuous $R$-symmetry appears. This
continuous $R$-symmetry is somewhat problematic since it forbids
gaugino Majorana masses and at least for the case of the gluino we
might have experimental evidence that its mass cannot vanish. Thus the
$R$-symmetry has to be broken. Since only a spontaneous breakdown of
this symmetry is acceptable, this then would lead to an embarrassing
Goldstone boson. Actually in our case it will be an axion since the
$R$-symmetry is anomalous[7]. This then tells us that this spontaneous
breakdown cannot happen at an energy scale like 100GeV.
 The breakdown scale of the R-symmetry has to be larger to make
the axion invisible[8], i.e. a breakdown scale of something like
$10^{10}$   to $10^{11}$GeV.
In a simple way this can, however, only be realized if also
the supersymmetry breakdown scale $M_S$ is large. Now remember that the
 splitting of the multiplets is given by
$\Delta m^2\sim gM_S^2$ where  $g$  is the coupling to
the goldstino. We thus need small couplings to have the
supersymmetric partners of quarks and leptons in the TeV-range to
provide us with a physical cutoff that stabilizes $M_W$. These
couplings have to be really small, compare them e.g. with the
gravitational coupling constant $\kappa$. We have
$$\delta m\sim \kappa M_S^2=M_S^2/M_P\eqno(3.11)$$
which is in the TeV-region for $M_S  = 10^{11}$GeV. Actually if we assume
that all particles couple universally to gravity our requirement of
the mass splittings implies $M_S$ to be approximately
$10^{11}$GeV. It is thus natural to
assume that the small coupling required from our discussion about
$R$-symmetry is actually the gravitational coupling constant[17].

We consider this as a hint to include gravity in our framework. This
will lead us to the local version of supersymmetry which includes
gravity automatically. It will turn out that such considerations
avoid some problems connected with the breakdown of global supersymmetry
 and
their desastrous consequences for model building. We shall not discuss
this here in detail and refer the reader to ref. [5] for a review.

   Local supersymmetry[9] will also resolve the paradox concerning the
nonzero cosmological constant in models
of spontaneously broken global supersymmetry.
We shall see that one can have $E_{\rm vac}=0$ in models of spontaneously
broken local supersymmetry.

\bigskip\bigskip

\line{\bf 4. Supergravity\hfill}
\medskip

In local supersymmetry the transformation parameter is
no longer constant but depends on space-time[10]. We have already
acquired some
experience in the framework of gauge symmetries:
 the local form of ordinary global
symmetries; and for supersymmetry we proceed in the same way. In
usual symmetries we had a scalar transformation parameter $\Lambda$. The
requirement of local invariance then leads to the introduction of a
gauge field  $A_\mu$ with transformation
property $\delta A_\mu=\partial_\mu\Lambda$. In supersymmetry we
have a spinorial parameter $\epsilon_\alpha$.
Local supersymmetry then requires the
introduction of a gauge particle $\Psi_{\mu\alpha}$  (the gravitino) with
transformation property
$\delta\Psi_{\mu\alpha}=\partial_\mu\epsilon_\alpha(x)$.
Thus the gauge particle of local supersymmetry is a spin 3/2 particle
and for reasons that will become clear in a moment it is called the
gravitino. These statements can also be made plausible when we
discuss the Higgs effect. In ordinary global symmetries a spontaneous
breakdown implied the existence of Goldstone bosons. In the local
version these bosons then supply the gauge bosons with the missing
degrees of freedom to make them massive. In supersymmetry the
goldstone particle is a spin 1/2 fermion. This then can provide the
two degrees of freedom in the transition of a massless to massive
spin 3/2 particle: the super-Higgs effect.

   The next point to discuss shows a conceptual difference between
ordinary symmetries and supersymmetry. While in ordinary theories it
was sufficient for the local symmetry to introduce a spin 1 gauge
boson in supersymmetry this is not the case. The gauge particle is a
spin 3/2 fermion and supersymmetry requires a bosonic partner. The
construction of local supersymmetry has shown that this partner is a
spin 2 boson that has to have all the properties of the graviton.
This then implies that local supersymmetry necessarily includes
gravity. We could have guessed that already from the algebra
$$[\epsilon(x)Q,\bar Q\bar\epsilon(x)]=
2\epsilon(x)\sigma_\mu\bar\epsilon(x)P^\mu.\eqno(4.1)$$
On the right hand side we have a space-time translation that
differs from point to point, a general coordinate transformation.

We have now to discuss explicit Lagrangians containing chiral matter
and gauge fields coupled to the (2,${3\over 2}$)-supergravity multiplet. In
general this requires a lot of tedious calculations which I shall not
repeat here. Also the general form of the Lagrangian is quite lengthy
and I refer to the literature for the complete expression[11].
I will instead concentrate on an analysis of the scalar
potential of these theories which we need for our further discussion.

   Remember that in the global case the most general Lagrangian was
defined by three functions of the superfields: the gauge kinetic
terms $W^2$, the matter field kinetic
terms $S(\phi^*\exp(gV)\phi)$ and the
superpotential $g(\phi)$. In the local case the most general action can
be defined by $f_{\alpha\beta}(\phi)W^\alpha W^\beta$
(with indices $\alpha$, $\beta$ labeling the adjoint
representation of the gauge group) and the K\"ahler potential
$$G=3\log\left(-{S\over 3}\right)-\log(\vert g\vert^2).\eqno(4.2)$$
The kinetic terms of the scalar particles $z_i$ are then given by
$$G^i_jD_\mu z_iD^\mu z^{j*}=
{{\partial^2G}\over{\partial z_i\partial z^{j*}}}
D_\mu z_i D^\mu z^{j*}\eqno(4.3)$$
where $z_i$ is the lowest component of a chiral superfield $\phi_i$. The
scalar potential reads
$$V=-\exp(-G)[3+G_k(G^{-1})^k_lG^l]+{1\over 2}f^{-1}_{\alpha\beta}
D^\alpha D^\beta.\eqno(4.4)$$
In these lectures I will use what is called minimal kinetic terms
$$G^i_j=-\delta^i_j.\eqno(4.5)$$
This simplifies all our formulas considerably and allows us
nonetheless to see all the essential properties of the potential. The
K\"ahler potential can therefore be written as
$$G=-{{z_iz^{i*}}\over M^2}-\log{{\vert g\vert^2}\over M^6}\eqno(4.6)$$
where we have explicitly written out the mass scale  $M$  related to
the gravitational coupling constant  $\kappa$:
$$M={1\over\kappa}={{M_{\rm Planck}}\over{\sqrt{8\pi}}}\approx
2.4\times 10^{18}{\rm GeV}.\eqno(4.7)$$
The first derivative of the K\"ahler potential is then given by
$$G^i=-{z^{i*}\over M^2}-{{g^i(z_i)}\over g(z)}\eqno(4.8)$$
and we can rewrite the potential in terms of the superpotential $g(z)$
as
$$V=\exp\left({{z_iz^{i*}}\over M^2}\right)\left[\left\vert g^i+{z^{i*}\over
M^2}g\right\vert^2-{3\over M^2}\vert g\vert^2\right].\eqno(4.9)$$
Contrary to the case of global supersymmetry the potential is no
longer semipositive definite. I still have to tell you under which
conditions supersymmetry is spontaneously broken. As in the global
case this breakdown is signaled by a vacuum expectation value of an
auxiliary field. There we had  the auxiliary field $F$ given
as the derivative of the superpotential; here we have an additional term
$$F^i=g^i+{z^{i*}\over M^2}g\eqno(4.10)$$
where in the limit  $M\rightarrow\infty$ we
recover the global result. Supergravity
is now spontaneously broken if and only if an auxiliary field
receives a vev. The supergravity breakdown scale is found to be
$$M_S^2=<F>\exp\left({{z_iz^{i*}}\over M^2}\right).\eqno(4.11)$$
Observe that the vacuum energy is no longer an order parameter. We
can have unbroken supergravity with $E_{\rm vac}<0$ (Anti de Sitter) or
$E_{\rm vac}= 0$ (Poincare supersymmetry) and
$E_{\rm vac}>0$ always implies broken supergravity. The
most important observation is, however, that we can have broken
supergravity with vanishing vacuum energy (cosmological constant), a
situation that could not occur in  the framework of
global supersymmetry. Here we need
$$\sum_i F^i F_{i}^*={3\over M^2}\vert g\vert^2\eqno(4.12)$$
and we will assume this to be fulfilled. In all cases I know
of this is an ad hoc adjustment of the cosmological constant to
zero. If (4.12) is fulfilled and if $M_S\not=0$ the gravitino becomes massive
through the super-Higgs effect
$$m_{3/2}=M\exp(-G/2)={g\over M^2}\exp\left({{z_iz^{i*}}\over M^2}
\right)\eqno(4.13)$$
and we therefore have the relation
$$m_{3/2}={M_S^2\over {{\sqrt 3}M}}\eqno(4.14)$$
valid in the case of vanishing cosmological constant.

   Let us now discuss some simple specific models with
spontaneous supersymmetry
breakdown. As a warm-up example consider one field  $z$  and  a
constant superpotential $g = m^3$. The potential is then given by
$$V=m^6\exp\left({{zz^*}\over M^2}\right)\left[
{{\vert z\vert^2}\over M^4}-{3\over M^2}\right]\eqno(4.15)$$
which has stationary points at $z = 0$ and
$\vert z\vert={\sqrt 2}M$. At $z = 0$
supersymmetry is unbroken but this is a local maximum of the potential.
The minima
with broken supersymmetry and $E_{\rm vac}<0$ are at $z =\pm{\sqrt 2}M$.

   Next we want to give an example with broken supersymmetry and
$E_{\rm vac}= 0$. We consider a superpotential
$$g(z)=m^2(z+\beta)\eqno(4.16)$$
A nonvanishing vev of
$$F={{\partial g}\over {\partial z}}+{z^*\over M^2}g=
m^2\left(1+{{z^*(z+\beta)}\over M^2}\right)\eqno(4.17)$$
would signal a spontaneous breakdown of supergravity. The equation
$$M^2+zz^*+z^*\beta=0\eqno(4.18)$$
has the solutions
$$z=-{\beta\over 2}\pm{1\over 2}\sqrt{\beta^2-4M^2}.\eqno(4.19)$$
Since (4.18) only allows real solutions (we assume $\beta$ to be real)
(4.19) implies that supersymmetry is broken as long
as  $\beta< 2M$. Thus
we can arrange for a supersymmetry breakdown but we still have the
annoying task to fine tune the vacuum energy. Let us therefore first
consider the case $\beta = 0$ in which the potential is proportional to
$$(M^2+\vert z\vert^2)^2-3M^2\vert z\vert^2\eqno(4.20)$$
which is positive definite with minimum at $z = 0$. Increasing  $\beta$
implies decreasing the vacuum energy and also  $z$  aquires a
nonvanishing vev. We can now increase $\beta$ until the potential just
touches zero. This is found to happen at
$\beta=(2 -{\sqrt 3})M$ with a vev of $({\sqrt 3} - 1)M$
for the $z$-field. The potential is semipositive definite
with $E_{\rm vac}= 0$ and, since
$\vert\beta\vert<2M$,  supersymmetry is broken and we have
found the desired example. The super-Higgs effect occurs. The
gravitino swallows the fermion in the chiral superfield and has a mass
$$m_{3/2}={m^2\over M}\exp\left( {{({\sqrt 3}-1)^2}\over 2}\right)\eqno(4.21)$$
and the two remaining scalars have masses
$$\eqalign{m_1^2&=2{\sqrt 3}m_{3/2}^2\cr
           m_2^2&=2(2-{\sqrt 3})m_{3/2}^2.\cr}\eqno(4.22)$$
Supersymmetry is broken and $E_{\rm vac}$ remains zero. Observe that such a
situation is not possible in the framework of global supersymmetry.
Observe also, that in the present example we had to perform an
explicit fine-tuning to obtain $E_{\rm vac}   = 0$.

   Before closing this chapter let us discuss two more examples of interest.
The first is supersymmetry breakdown through
gaugino-condensation. Consider a {\it pure} supersymmetric gauge theory,
just a gauge theory with fermions (the gauginos) in the adjoint
representations of the gauge group. Such a theory is asymptotically
free, the gauge coupling becomes strong at small energies and we
assume, in analogy to QCD, that this leads to confinement and that
gaugino bilinears condense. For a detailed discussion see ref.[12]. To
see whether this leads to supersymmetry breakdown we have to consider
the auxiliary fields of supergravity including the gaugino bilinears
$$F_i=\exp(-G/2)(G^{-1})_i^jG_j+{1\over 4}f_{\alpha\beta k}
(G^{-1})_i^k(\lambda^\alpha\lambda^\beta)+\ldots\eqno(4.23)$$
where $\lambda^\alpha$ are the gauginos,
$f_{\alpha\beta}$ the socalled gauge-kinetic function
that multiplies $W^\alpha W^\beta$    and
$f_{\alpha\beta k}=\partial f_{\alpha\beta}/\partial z^k$. A nontrivial vev
$<\lambda\lambda>\not= 0$
thus breaks supersymmetry provided that the gauge kinetic function is
nontrivial[13]. The supersymmetry breakdown scale is given by
$$M^2_S\sim {{<\lambda\lambda>}\over M}\eqno(4.24)$$
leading to a gravitino mass of order
$<\lambda\lambda>/M^2$. Observe that the
value of $M_S$ in (4.24) vanishes in the
global limit $M\rightarrow\infty$. Models in
which supersymmetry breakdown is induced by gaugino condensation have
recently attracted revived attention because
of their appearance in the low energy
limit of string theories. They are also interesting because of the
fact that for a nontrival  $f$ the value of the gauge coupling
constant $g^2\sim 1/f$  is a dynamical parameter. In string theories it
is related to the vev of the dilaton   field[14].

   Up to now we have for the sake of simplicity only discussed models
with minimal kinetic terms for the scalar fields. Models with
nonmininal kinetic terms can have interesting structure. Consider
e.g.
$$G=3\log(\phi+\phi^*)-\log\vert g\vert^2\eqno(4.25)$$
and take a constant superpotential. If you compute the potential as
given in (4.4) you will find that it vanishes identically. Nontheless
the quantity
$$e^{-G}={{\vert g\vert^2}\over {(\phi+\phi^*)^3}}\eqno(4.26)$$
does not vanish and supersymmetry is broken. Such
so-called no-scale
models[15] might also have applications in the low energy limit of string
theories.

\bigskip\bigskip

\line{\bf 5. Low energy supergravity models\hfill}
\medskip

   As we discussed in chapter 3 we should consider models that
consist of two sectors: a hidden sector and an observable sector
which are only coupled weakly through gravitational interactions. The
observable sector consists of the fields discussed in chapter 3 which
we will collectevely denote by $y_a$. The hidden sector is responsible
for the breakdown of supersymmetry at a scale $M_S\sim 10^{11}$GeV and leads to
a gravitino mass in the TeV region. Its fields will be denoted by $z_i$
and we choose a superpotential
$$\tilde g(z_i,y_a)=h(z_i)+g(y_a).\eqno(5.1)$$
Let us parametrize a general hidden sector by assuming that at the
minimum
$$\eqalign{<z_i>&=b_iM\cr
           <h >& =mM^2\cr
      <h_i>&=<\partial h/\partial z_i>=a_i^*mM\cr}\eqno(5.2)$$
while all abserable sector fields $y_a$ should have vanishing vev's. In
the example of last chapter we had e.g.
$b ={\sqrt 3}-1$. The potential is
given by
$$
V=\exp\left({{\vert z_i\vert^2+\vert y_a\vert^2}\over M^2}\right)
\left[\left\vert h_i+{{z_i^*\tilde g}\over M^2}\right\vert^2+
\left\vert g_a+{{y_a^*\tilde g}\over M^2}\right\vert^2-{3\over M^2}
\vert \tilde g \vert^2\right].\eqno(5.3)$$
The vacuum energy vanishes provided
that
$$\sum_i\vert a_i+b_i\vert^2=3\eqno(5.4)$$
and the gravitino mass is given by
$$m_{3/2}=\exp\left({1\over 2}\vert b_i\vert^2\right)m,\eqno(5.5)$$
thus $m$ sets the scale of the gravitino mass.
We furthermore define[16]
$$A=b_i^*(a_i+b_i)\eqno(5.6)$$
which will turn out to be an important parameter besides the gravitino
mass. In the previous example we had
$A = 3-\sqrt 3$. The potential given
in (5.3) is complicated but we have $m \ll M$  and we can simplify the
expressions enormously by neglecting subleading terms. Formally this
means that we take the limit
$M\rightarrow\infty$ keeping, however, $m_{3/2}$ fixed. We
then replace the hidden sector fields by there vev's and obtain the
following potential for the observable sector fields
$$V=\left\vert{{\partial g}\over {\partial y_a}}\right\vert^2+
m^2_{3/2}\left\vert y_a\right\vert^2+m_{3/2}
\left[y_a{{\partial g}\over{\partial y_a}}+(A-3)g +{\rm h.c.}\right].
\eqno(5.7)$$
Thus the spontaneous breakdown of supergravity in the hidden sector
manifests itself as explicit breakdown of global supersymmetry in the
low energy limit of the observable sector. The first term in (5.7) is
the usual potential of a globally supersymmetric theory while the
other terms are soft breaking terms.

  The second term gives universal scalar masses to all the partners
of quarks and leptons. The supertrace formula is here given in
general by[11]
$${\rm STr}M^2=2(N-1)m^2_{3/2}\eqno(5.8)$$
where  $N$  is the number of chiral superfields. This avoids the mass
relations obtained in the globally supersymmetric models and its
desastrous consequences for model building. The universality property
of the mass terms is needed to ensure the absence of flavour changing
neutral currents. It appears here because of the choice of minimal
kinetic terms for the scalar fields.

   The term $(A-3)g$ is of equal importance since it breaks all
$R$-symmetries of the model. This implies that there are no problems
with potential axions and that also gaugino Majorana masses are
allowed (recall our discussion in chapter 3). This breakdown of
$R$-symmetry is a direct consequence of the coupling to gravity.

  One more technical remark. In general we will deal with a
superpotential $g = g_3  + g_2$ where $g_3$ denotes the trilinear and $g_2$
the bilinear terms. The last term in (5.7) then reads
$Am_{3/2}g_3 +(A-1)m_{3/2}g_2$. Apart grom the gaugino mass
$m_0$  we find that $m_{3/2}$ (which sets the scale for
the soft scalar masses) and
$A$  are the important parameters parametrizing the effects of
supersymmetry breakdown in this class of models. In some cases one can
also consider a new parameter $B$ as the coefficient of the bilinear
terms in the superpotential. In the simplest example $B=A-1$, but this need not
be the case in general.

A remark about the mechanism of SUSY-breakdown is in order here. The example
of one scalar field with superpotential (4.16) should, of course, only
be considered as a toy example and
existence proof for such a mechanism. The true
mechanism of SUSY-breakdown will certainly look different, already because
of the fact that the scale of $10^{11}$ GeV has to be put in by hand.
Nowadays the most discussed mechanism for SUSY-breakdown is based on the
mechanism of gaugino condensation[12]. Here the SUSY breakdown scale
can be understood dynamically as a consequence of a new strong gauge
coupling, in a similar way as we can understand the mass of the proton
through the scale of QCD. One should also remark that a model based on
SUSY breakdown through gaugino condensation initiated the construction
of hidden sector models based on broken supergravity[17]. Later it was found
that such a mechanism fits very well in the framework of models derived
from heterotic string theory[18]. Therefore this mechanism of
SUSY-breakdown is very popular at present.

 Let us now discuss the superpotential
$$g=\mu H\bar H+g_EHL\bar E+g_DHQ\bar D+g_U\bar HQ\bar U.\eqno(5.9)$$
The parameter $\mu$  has to be different from zero since otherwise we
would have problems with a light higgsino (the supersymmetric partner
of the Higgs-scalar) or axions. The value of $\mu$ is not directly related
to the supersymmetry breakdown scale but one can construct models
[19] where
$\mu$ is related to $m_{3/2}$ and we shall assume
that also $\mu$ is in the TeV range.

  Let us now address the question of $SU(2)\times U(1)$ breakdown. We
have two Higgs multiplets and members of both have to receive
nonvanishing vev's to give masses to all quarks and leptons,
according to (5.9). The relevant part of the Higgs potential reads[20]
$$V=m^2_1\vert h\vert^2+m_2^2\vert \bar h\vert^2+m_3^2(h\bar h+h^*\bar h^*)+
{{g_1^2+g_2^2}\over 8}(\vert h\vert^2-\vert \bar h\vert^2)^2\eqno(5.10)$$
where the last term corresponds to the $SU(2)\times U(1)$  $D$-term and $g_2$
and $g_1$  denote the respective coupling constants. From (5.7) and (5.9)
we obtain
$$\eqalign{m_1^2&=m_2^2=m_{3/2}^2+\mu^2\cr
m_3^2&=-B\mu m_{3/2}\cr
B&=A-1\cr}\eqno(5.11)$$
The potential consists of quadratic and quartic terms. The quartic
terms have a positive coefficient such that the potential at infinity
is well behaved, with the exception, however, of a flat
direction for   $\vert h\vert=\vert\bar h\vert$.
 To have the potential bounded from
below we therefore have to impose a constraint on the coefficients of
the quadratic terms
$$m_1^2+m_2^2\geq 2\vert m_3^2\vert.\eqno(5.12)$$
Next we have to discuss the requirement of $SU(2)\times U(1)$ breakdown.
Since there are no trilinear terms in (5.10) a stationary point at
$h=\bar h = 0$ has to be unstable, i.e. the mass matrix at this point has to
have a negative eigenvalue. The requirement for a nontrivial
$SU(2)\times U(1)$ breaking absolute minimum is therefore
$$\vert m_3\vert^4\geq m_1^2m_2^2.\eqno(5.13)$$
With the input parameters (5.11) we observe now that the
constraints (5.12) and (5.13) can only be fulfilled in the limiting case
$$m_{3/2}^2+\mu^2=B\mu m_{3/2}\eqno(5.14)$$
i.e. at most we can arrive at a flat direction where $SU(2)\times U(1)$
breaking and nonbreaking minima are degenerate. We would then have to look
for radiative corrections to see whether $SU(2)\times U(1)$ breaking minima
can be reached at all within this approach. This is actually a nice
feature of the model. It tells us again that we have not put in
$SU(2)\times U(1)$ breaking by hand. Instead
this breakdown will be intrinsically
related to the supersymmetry breakdown and the dynamics of the model.
The investigation of this question involves a full treatment of the
evolution of the parameters of the model in the framework of the
renormalization group approach. Time does not permit us to discuss
that in detail. We refer the reader to some existing reviews that also
discuss the phenomenological properties of the model[5,21].

\bigskip\bigskip

\line{\bf 6. Grand unification\hfill}
\medskip

Again I assume that the reader is familiar with the general idea of grand
unification[22].
We shall  concentrate here on those special points that are
important in the supersymmetric case.
 This concerns the scale $M_X$, a discussion of the
superpotential, the question of the triplet-doublet splitting and
proton decay via dimension 5 operators. We shall exclusively stay
within the $SU(5)$ framework, with $\bar 5 + 10$ for a quark-lepton family.

In this first chapter on supersymmetric grand unification we give the basic
structure of these theories. A more careful discussion of the models
including the results of recent precision measurements will be given
in the next chapter.
 If we very
roughly assume a value of $\alpha_3\sim 0.11$  and
$\alpha\approx 1/129$ at a scale of 100 GeV we obtain in the nonsupersymmetric
model a scale $M_X$  of approximately $5\times 10^{14}$ GeV
and desastrous proton
decay. The supersymmetric model, however, has more light particles
and as such the evolution of coupling constants changes[23]. The most
important contribution comes from the gauginos implying a slow-down
of the evolution. As a result we observe a larger $M_X\sim 2\times 10^{16}$ GeV
roughly 60 times larger than in the corresponding nonsupersymmetric
model. Since proton decay is suppressed with the fourth inverse power
of $M_x$  there are no problems with proton stability in the
supersymmetric $SU(5)$ model.
For a long time the experimental uncertainties concerning the value of
the gauge coupling constants did not allow a distinction between the
supersymmetric and nonsupersymmetric models.
But more recently this
situation has changed.
A precision analysis of
electroweak data indicated that the supersymmetric model
(with two Higgs doublets and a supersymmetry breakdown
scale in the TeV-region) gives, in
constrast to nonsupersymmetric $SU(5)$ the correct prediction for
$\sin^2\theta_W(M_Z)$ [24]. We shall discuss these questions
in the next section. First we would like to present some basic facts
of supersymmetric grand unified theories.

   Let us here next examine the superpotential and the question of $SU(5)$
breakdown. We denote the quark superfields $X_i$(10), $Y_i(\bar 5)$
$i =1,2,3$ and the Higgs superfields $H(5)$, $\bar H(\bar 5)$
and $\Phi(24)$. The
superpotential can then be writen as
$$g=g_{ij}X_iX_jH+h_{ij}X_iY_j\bar H+\lambda_1H\Phi\bar H+
\lambda_2\Phi^3+M\Phi^2+M^\prime H\bar H\eqno(6.1)$$
where $g_{ij}$ determines the masses of up-type
quarks and $h_{ij}$  those of
down-type quarks and leptons. The discussion of
the spontaneous breakdown of $SU(5)$ is
similar to the one in nonsupersymmetric $SU(5)$ models. The auxiliary
fields read
$$\eqalign{-F_\Phi^*&=\lambda_1H\bar H+3\lambda_2\Phi^2+2M\Phi\cr
           -F_H^*&=\lambda\Phi\bar H+M^\prime\bar H+g_{ij}X_iX_j\cr
           -F_{\bar H}^*&=\lambda_1\Phi H+M^\prime H+h_{ij}X_iY_j\cr}
\eqno(6.2)$$
and a minimum with $SU(5)$ broken to
$SU(3)\times SU(2)\times U(1)$ can be found
with $<H>=<\bar H>=<X_i>=<Y_i>=0$,
$$<\Phi>=v\pmatrix{1&0&0&          0&0\cr
                   0&1&0&          0&0\cr
                   0&0&1&          0&0\cr
                   0&0&0&-{3\over 2}&0\cr
                   0&0&0&0&-{3\over 2}\cr}\eqno(6.3)$$
and vanishing vacuum energy. Since we have not discussed here the
breakdown of supersymmetry there are degenerate minima with gauge
group $SU(5)$ and $SU(4)\times U(1)$. Also the breakdown of
$SU(2)\times U(1)$ has
finally to be induced by the effects of supersymmetry breakdown along
the lines discussed in the last chapter.

 Again a fine-tuning has to be performed to keep the Higgs-doublets
light. Here it amounts to
$$M^\prime ={3\over 2}v\lambda_1.\eqno(6.4)$$
This is similar to the nonsupersymmetric case but here we could argue
that the fine-tuning concerns only parameters in the superpotential
and is therefore not  disturbed by radiative corrections. If we now
would be able to find a reason why (6.4) should be valid at tree
level we could claim to have solved the fine-tuning problem. There
have been several interesting attempts in this direction. As a first
we discuss the mechanism of a sliding singlet[25]. Take a gauge singlet
superfield  $Z$  and add a term
$\lambda HZ\bar H$ to the superpotential. The $H$
auxiliary field reads now
$$-F_H^*=\bar H(\lambda_1\Phi+\lambda Z+M^\prime).\eqno(6.5)$$
In the full theory, including supersymmetry breakdown, the doublet
component of  the scalar of $\bar H$ should receive a vev (in contrast to
the $SU(3)$-triplet component). The vev of $Z$ is undetermined and it
can adjust its vev to have $F =0$   for the doublet component, thus it
slides to make
$$-{3\over 2}\lambda_1 v+\lambda z +M^\prime=0\eqno(6.6)$$
and the Higgs-doublet remains light. This looks nice, but also this
mechanism has some problems. We do not understand why the
allowed $Z^2$  and $Z^3$
terms are absent and also we cannot rule out the possibility that the
absolute minimum of the potential occurs for large vev's of both the
triplet and the doublet. Moreover, there are usually problems with a
small supersymmetry breakdown scale in the presence of light singlets[26].

   A second mechanism to be discussed here is the one of the missing partner.
$H$ and $\bar H$ contain $(3,1) + (\bar 3,1)$ and
$(1,2) + (1,\bar 2)$ of $SU(3)$ and $SU(2)$
respectively. Try to find now a new representation which only
contains a (3,1) but not a (1,2). The former could then pair up with
the $(\bar 3,1)$ in  $\bar H$  while $(1,\bar 2)$ would remain massless. A
simple example[27] is a 50 of $SU(5)$. It decomposes with respect to
$SU(3)\times SU(2)$ as $(\bar 6,1)+(8,2)+(1,1)+(3,2)+(6,3)+(\bar 3,1)$ and
as a cross term in the superpotential we could imagine
$50\times 5\times 75$ with
$75=(1,1)+(3,1)+(3,2)+(\bar 3,1)+(\bar 3,2)+(\bar 6,2)+(6,2)+(8,1)+(8,3)$.
Fortunately a vev of 75 can break $SU(5)$ to
$SU(3)\times SU(2)\times U(1)$ thus
avoiding the presence of $\Phi$ in (6.1).
Instead we choose now for the superpotential
$$\eqalign{g&=\lambda 75\times 75\times 75+M75\times 75+
\lambda_1 50\times 75\times\bar {50}\cr
&+\lambda_2 50\times 75\times 5+\lambda_3 \bar {50}\times 75\times\bar 5+
\bar M 50\times\bar {50}\cr}\eqno(6.7)$$
and as a mass matrix for the triplets we obtain
$$\pmatrix{0&\lambda_2 v\cr
          \lambda_3 v&\bar M\cr}\eqno(6.8)$$
(where  $v$  is the vev of 75), while the doublets remain light. Of
course, one still has to explain why we have omitted a direct
$5\times\bar 5$
mass term in (6.7) and the question of a complete solution of the
fine tuning problem remains open.

   We had seen at the beginning of this chapter that  $M_X$ is quite
large in supersymmetric grand unified models and that therefore
proton decay via gauge boson exchange is sufficiently suppressed.
This, however, is not the last word about proton decay in
supersymmetric grand
unified models. Remember, that in the supersymmetric version of
the standard model we already had to suppress proton decay via
dimension-4 operators by introducing an $R$-symmetry (see chapter 3).
Here we have to worry about proton decay via dimension five operators[28]
leading to proton decay via the exchange of fermionic
supersymmetric partners. The first step
couples two fermions to two bosons (therefore the name dimension-5
operator) and has a propagator suppression of $1/M_X$  and the second
step involves only light particles. Instead of $1/M_X^2$ in the amplitude
we have now $1/M_X M_W$ and there is a potential danger of fast proton decay.
A careful investigation of the dimension 5-operators has therefore
to be performed. Out of the possible terms we need only consider those
which are invariant under the $R$-symmetry discussed earlier and these
are the two $F$-terms $(QQQL)_F$
 and $(\bar U\bar U\bar D\bar E)_F$ . The latter reads in components
$$\bar U_{ia}\bar U_{jb}\bar D_{kc}\bar E_l\epsilon^{abc}\eqno(6.9)$$
where $a$, $b$, $c$ are $SU(3)$ indices and $i$,
$j$, $k$, $l$ are generation indices. All
fields above are scalar superfields and should obey Bose-statistics.
The two  $\bar U$'s are antisymmetrized in $a$ and $b$ and
therefore  $i\not= j$  and
one of the $\bar U$'s has to come from the second generation. Since the
charmed quark is heavier than the proton the
presence of the term in (6.9) does not constitute a
problem. The other possibility reads
$$Q^a_{ir}Q^b_{js}Q^c_{kt}L_{lu}\epsilon_{abc}\epsilon^{rs}
\epsilon^{tu}\eqno(6.10)$$
where $r$, $s$, $t$, $u$ are $SU(2)$-indices.
Here we can have $i=j=1$ but then we
need $k=2$ which leads to
$${c\choose s}_t{\nu\choose e}_u\epsilon^{tu}\eqno(6.11)$$
thus $ce$  or  $s\nu$. Proton decay therefore is only possible with the
$(uds\nu)_F$   operator
The dominant decay mode is proton to $K^+$ and antineutrino, a quite
unique prediction of supersymmetric grand unified models. The rate is
faster than the one from dimension-6 operators but it is not
desastrously fast since $p\rightarrow K^+\bar\nu$
involves Yukawa couplings instead of gauge couplings in the process
with dimension-6 operators.
  At the moment $p\rightarrow K^+\bar\nu$
seems to be at the border of observability and further experimental
results are eagerly awaited.
\bigskip
\bigskip

\line{\bf 7. Supersymmetric grand unification\hfill}
\medskip

Experimental findings give at the moment the following picture; with a top
quark mass between 150 and 200 GeV the strong coupling constant
$\alpha_s(M_Z)=0.12\pm0.01$ and $\alpha_{\rm em}(M_Z)=1/128$ the weak mixing
angle is $\sin^2\theta_W(M_Z)=0.2316\pm 0.0003$. This leads to
gauge coupling unification at a scale $M_X=2\times 10^{16}$GeV with
$\alpha_{\rm GUT}\sim 1/26$ in the minimal supersymmetric extension of
the standard model, provided the mass scale $M_{\rm SUSY}$ of the
supersymmetric partners is between 100 Gev and 10 TeV.
There is a correlation between $\alpha_s$ and $M_{\rm SUSY}$: large
$\alpha_s$ corresponds to small $M_{\rm SUSY}$.

We have now to take a closer look at the definition and the role of
$M_{\rm SUSY}$. It is understood that between $M_Z$ and $M_{\rm SUSY}$
one should use the renormalization group equations of the standard
model while above $M_{\rm SUSY}$ the evolution equations of the
supersymmetric extension of the standard model should be applied.
If we consider e.g. a model where all the supersymmetric partners like the
gauginos, the higgsinos, the squarks and the sleptons are degenerate
with mass $m$, then $M_{\rm SUSY}=m$; this in fact would then mean,
that $M_{\rm SUSY}=m\geq M_Z$.

A more realistic spectrum of supersymmetric partners, however,
might look different. We e.g. expect in general, that
the gluino is heavier than the photino or that the squarks are heavier than the
sleptons; in any case one would expect a nondegenerate spectrum.
Some averaging proceedure should then be performed. It turned out that
strange things happen in this proceedure. It was
observed that even
with nondegenerate partners {\it all in mass above} $M_Z$ the effective scale
$M_{\rm SUSY}$ can become {\it smaller} than $M_Z$. We shall therefore
(following ref.[30]) call this effective scale $T_{\rm SUSY}$
and still keep the notation $M_{\rm SUSY}$ for the physical mass scale
of supersymmetric partners.

This effect of the averaging proceedure for a nondegenerate spectrum
has been explained in ref. [29]. Let us here follow
this discussion and use the evolution equations at the one-loop
level. The qualitative features are valid also if we include the
two-loop contribution, but the formulae become too complicated to
be discussed here. Here we obtain the following relation:
$$19\log\left({{T_{\rm SUSY}}\over{M_Z}}\right)=
 -25\log\left({{M_1}\over{M_Z}}\right)
+100\log\left({{M_2}\over{M_Z}}\right)
 -56\log\left({{M_3}\over{M_Z}}\right),\eqno(7.1)$$
where $M_1$, $M_2$ and $M_3$ in some way represent the average mass of
particles with $U(1)$, $SU(2)$ and $SU(3)$
quantum numbers, respectively[29].
At the moment it is not necessary to understand these masses in detail;
we will shortly give a more detailed explanation. It is important to
realize  first that in fact the whole spectrum can be described by
{\it one} effective scale $T_{\rm SUSY}$ that represents all information
about these threshold corrections for the supersymmetric particles.
Secondly we observe that the right hand side of (7.1) contains positive
as well as negative signs. And here we now understand the strange
behaviour mentioned above: if we increase the mass of the gluino (contributing
only to $M_3$) while keeping all other masses fixed we lower the
effective scale $T_{\rm SUSY}$. This also makes clear that it is
possible to have $T_{\rm SUSY}<M_Z$. The threshold effects that take
place above $M_Z$ and which might come from a complicated spectrum
can be summarized with this one effective scale $T_{\rm SUSY}$.

Let us now examine more closely the effect of the various particles on
$T_{\rm SUSY}$:
$$\eqalign{-19\log\left({{T_{\rm SUSY}}\over{M_Z}}\right)&=
3\log\left({{M_{\rm squarks}}\over{M_Z}}\right)
+28\log\left({{M_{\rm gluino}}\over{M_Z}}\right)\cr
&-3\log\left({{M_{\rm slepton}}\over{M_Z}}\right)
-32\log\left({{M_{\rm wino}}\over{M_Z}}\right)\cr
&-12\log\left({{M_{\rm higgsino}}\over{M_Z}}\right)
-3\log\left({{M_{\rm Higgs}}\over{M_Z}}\right),\cr}\eqno(7.2)$$
which allows you to compute $T_{\rm SUSY}$, once you know the masses
of the particles in
the supersymmetric standard model. It is clear that even with
all the supersymmetric partners heavy, still $T_{\rm SUSY}$ might be
small. Observe also, that the contribution from squarks and
sleptons cancel if they are degenerate. The terms with the gauginos have
quite large coefficients. If one considers models with a universal
gaugino mass at the large scale, there is also the tendency that the
winos partially cancel a big gluino contribution. In general,
however, threshold corrections due to the nondegeneracy of the
supersymmetric spectrum are quite important. This remains true after the
inclusion of two-loop effect in the evolution equations which we have not
discussed here. An account of the difference between one- and
two-loop results on these questions can be found in [31].

Let us now turn to the question of fermion masses in grand unified models.
 We remember that a discussion of the
quark and lepton masses in the standard model as well as its
supersymmetric extension usually consisted of the statement that one
has to adjust the Yukawa-couplings to obtain the correct spectrum.
In grand unified models, however, due to the larger gauge symmetry we also
have to consider {\it Yukawa coupling unification}. In our example
based on the group $SU(5)$ we have, according to equation (6.1), only
one Yukawa coupling for the charged leptons and the
down quarks, as long as we assume that they obtain their mass
through the vev of the same Higgs-scalar. The complete fermion
mass matrix is very complicated, and in order to understand it
completely one would most probably need more than just one of
these scalars. It is, however, tempting to assume that for the
heaviest generation just one scalar is responsible for b-quark and
$\tau$-lepton mass. This then implies $h_{\tau}(M_X)=h_b(M_X)$
for the b- and $\tau$-Yukawa-couplings at the GUT-scale.
Of course, at low energies, $h_\tau$ and $h_b$ differ because
of the renormalization effects. The one-loop
equations for the Yukawa-couplings are given by
(assuming $h_t\gg h_b$, $h_\tau$):
$${\tilde\mu{\partial\over{\partial\tilde\mu}}h_{\tau}=
-{1\over {8\pi^2}}h_{\tau}\left({3\over 2}g_2^2+{3\over 2}g_1^2
\right)}\eqno(7.3)$$
$$\tilde\mu{\partial\over{\partial\tilde\mu}}h_{b}=
-{1\over {8\pi^2}}h_{b}\left({8\over 3}g_3^2+
{3\over 2}g_2^2+{7\over 18}g_1^2\right)
+{1\over {16\pi^2}}h_t^2h_b\eqno(7.4)$$
$$\tilde\mu{\partial\over{\partial\tilde\mu}}h_{t}=
-{1\over {8\pi^2}}h_{t}\left({8\over 3}g_3^2+
{3\over 2}g_2^2+{13\over 18}g_1^2\right)
+{3\over {8\pi^2}}h_t^3\eqno(7.5)$$
in the notation of chapter 5. Since the b-quark has strong interactions
in contrast to the $\tau$-lepton $h_b$ evolves faster than $h_\tau$
giving rise to a larger b-mass at low energies in agreement with
experimental results. This is a well known result and it was considered
a great success that the $m_b/m_\tau$ ratio could be explained by this
fact[32].
In [33] it was pointed out, that for a large value
of $h_t$ (comparable in size to the gauge coupling constants) its effects
could be quite important. This comes from the last term in (7.4) with the
opposite sign, thus reducing the  $m_b/m_\tau$ ratio.
This ratio thus depends strongly on $\alpha_s$ and $h_t$. For a long
time $\alpha_s$ was so poorly known that no conclusion could be drawn from
these facts. With the more precise value of $\alpha_s$ and the knowledge
of the $m_b/m_\tau$ ratio now, however, we can obtain information on the size
of the top-quark Yukawa coupling $h_t$ [34].
This leads to the statement, that $h_t$ should be close to its
infrared quasi fixed-point [35] which is obtained in case of a
vanishing right hand side of (7.5), thus with
$Y_t=h_t^2/4\pi$
$$8\alpha_s(m_t)\sim 9 Y_t(m_t),\eqno(7.6)$$
evaluated at the low energy scale, here chosen to be the mass of the
top-quark. This value of $h_t$ close to the infrared quasi fixed point
leads to rather large values of the top-quark mass:
$$m_t(m_t)=h_t(m_t) v \sin\beta,\eqno(7.7)$$
where $\tan\beta$ is the previously defined ratio of the vevs of the
two Higgs-fields.

Observe  that in models with radiative symmetry breakdown one has
$\tan\beta\geq 1$ and thus $m_t\geq 140$ GeV approximately.
The assumption of {\it Yukawa coupling unification} for the
$b - \tau$ system gives strong restrictions on $m_t$.
A detailed discussion of these and related questions
can be found in the literature[30]. Meanwhile the
direct experimental observation of the top quark has confirmed these
expectations.

As we discussed in the last chapter there can be constraints from proton
decay via dimension-5 operators. This process involves the
down-quark Yukawa coupling and given the d-quark mass we see that this
coupling is proportional to $\tan\beta$. The experimental results
might therefore lead to an upper bound[36] on $\tan\beta$,
but the exact value of this bound is still under debate[37].
If proton decay via dinension-5 operators is not found one might also
consider  models where some discrete symmetries[38,39]
(as alternatives to R-parity) prohibit this mechanism. In these cases we
would, however, expect new sources of lepton number violation.

An upper bound on $\tan\beta$
 might become important in those models based on an $SO(10)$ grand unified
gauge group where the heaviest generation receives a mass from a single
Higgs representation. There $h_t=h_b$ at $M_X$ and therefore
$\tan\beta\sim 60$.

The simplest supersymmetric grand unified model is thus consistent with the
value of $\sin^2\theta_W$. Given this success, we can then test more specific
models, like the assumption of Yukawa-coupling unification discussed
above. Another more specific scenario is the one based on the induced
radiative breakdown of $SU(2)\times U(1)$. Here we obtain strong
restrictions on the parameters[40], especially in models that
also exhibit Yukawa coupling unification. At the moment
we can just try and study the full parameter space of the model.
New data has then to decide which part of it might be selected.
A lot of work has been done in this field recently which we do not have
the time
 to present in detail.

We should, however, be aware of the fact that in all grand unified models
there are inherent uncertainties at the grand scale that we cannot
control. These are e.g. threshold corrections due to heavy particles.
While in minimal $SU(5)$ they are usually rather mild[29], they could
become quite important in more complicated models like those with
a 75-representation discussed earlier[41]. Other uncertainties
include heavy thresholds in the evolution of Yukawa-couplings, the presence of
nonminimal gauge kinetic terms or just a more general
set of boundary conditions for the soft breaking terms at the grand scale[42].

\bigskip
\bigskip
\line{\bf 8. String unification\hfill}
\medskip

This brings us to the central question: should we believe in the reality
of supersymmetric grand unified theories? After all some ten years ago
many people believed in normal grand unified theories. Then proton
decay was not found and now we also know that the coupling constants in
a nonsupersymmetric theory do not match at a single scale. Could
history repeat itself? Of course, we cannot answer this question.
Nonetheless it might be useful to keep this possibility in mind.
If the GUT idea were true, however, we could then
ask the question how well we can determine the grand unified scale $M_X$
with our present experimental knowledge. That seems to be easy: just take
the precisely known values of $\alpha_1$ and $\alpha_2$ at $M_Z$ and
then determine the value where they cross. This would give something
like $M_X\sim 2\times 10^{16}$GeV. But we cannot control heavy threshold
effects and they might strongly influence the value of $M_X$. In fact,
grand unified models with a complete description of the fermion mass
spectrum turn out to lead to a complicated spectrum of heavy particles
and significant heavy threshold effects
might be a genuine property of realistic grand unified models. Also
$M_X$ tends to be only two orders of magnitude smaller than the Planck scale.
How sure can we be that $M_X\ll M_{\rm Planck}$ since gravitational effects
might also influence $M_X$.

We cannot answer these questions at the moment and one way to proceed is
to compare SUSY-GUTs with alternative models. One of them is the
embedding of the supersymmetric standard model within the framework of
string theory, called {\it string unification}. Such theories contain one
fundamental scale $M_{\rm string}\approx 4\times 10^{17}$GeV related to the
Planck scale. Many heavy particles can act as sources for threshold
effects. There is usually a fixed relation between the gauge coupling
constants but they need not necessarily all coincide at a single scale.
The models in general do not contain a grand unified gauge group like
$SU(5)$ or $SO(10)$ although such groups might be present. This could
relieve somewhat the problem of splitting doublet and
triplet in grand unified $SU(5)$ since the Higgs-doublet does not neccesarily
have an $SU(5)$ partner. It also implies that Yukawa couplings like
$h_b$ and $h_\tau$ need not be equal at the grand scale.

Let us now examine string unification in more detail. At the tree level
the gauge couplings are determined by the vev of the
scalar dilaton field:
$$k_3g_3^2=k_2g_2^2=k_1g_1^2=g_{\rm string}^2\equiv g^2\eqno(8.1)$$
where the coefficients $k_i$ (the so-called Kac-Moody levels) are
rational numbers. One could now try to see which choice for the
$k_i$ leads to models consistent with observed values of the
coupling constants. From the experience with model building we
know that it is very hard to obtain realistic models with $k\not=1$
for nonabelian gauge groups and one would choose $k_3=k_2=1$ leaving
$k_3$ as a free parameter. In SUSY-GUTs the relation between the
coupling constants would be fixed, but $M_X$ would be the free parameter
while in the other approach $M_X$ is fixed through $M_{\rm string}$.
The usual normalization of the
$U(1)$ gauge coupling corresponds to $k_1=5/3$.

The evolution of coupling constants requires a loop-calculation and
apart from the usual evolution the gauge couplings become moduli-dependent
(i.e. a function of scalar fields $T_i$) and this can be understood as the
influence of heavy particles. In simple models such a functional
dependence can be estimated [43-46] while in more realistic models
such a calculation turns out to be quite complicated[47]. One can write
$${{16\pi^2}\over{g_a^2(\mu)}}=k_a{{16\pi^2}\over{g_{\rm string}}}
+b_a \log(M^2_{\rm string}/\mu^2)+\Delta_a,\eqno(8.2)$$
where in the simplest cases[45] the threshold
$\Delta\sim\log[{\rm Im}T(\eta(T))^4]$ is a function of one modulus $T$
which is related to the overall size of compactified space. $M$ is
as [44]
$$M_{\rm string}=(2\pi\alpha^\prime)^{-1/2}\exp[(1-\gamma)/2] 3^{-3/4}
\approx 0.527 g_{\rm string}\times 10^{18}{\rm GeV}.\eqno(8.3)$$
If we then assume $g_{\rm string}\sim 0.7$ for the correct size of
the coupling constant $M_{\rm string}$
 turns out to be a factor of 20 larger than $M_X$. If we now,
hypothetically take $M_{\rm string}$ as the unification scale and
assume that all the
particles lighter than $M_{\rm string}$ are those of the minimal supersymmetric
standard model we can determine $\sin^2\theta_W\approx 0.218$ which is in
conflict with the measured value. This calculation, however, is purely
academic, since such a string model might not exist. Usually such models
contain
more  particles below $M_{\rm string}$ and thresholds might be important. Just
consider $\Delta(T)$ in its simplest form with a value of
$T\approx 10^{16}$GeV: one could have effects big enough to reproduce
the correct value of $\sin^2\theta_W$[48].
Of course, also such a calculation
might be academic since this simple threshold function is valid only in
very simple models with unbroken $E_6$ gauge group. It indicates, however,
the potential importance of heavy thresholds. This is confirmed in
more realistic models; see ref. [47] for a detailed discussion.
One way to distinguish string unification and grand unification could
be related to the question of Yukawa couplings. While in many
grand unified models with a simple Higgs sector we expect also some
group theory relations between Yukawa-couplings (like e.g.
$h_b=h_\tau$), this needs not necessarily be the case in string theory.
We have to see how the experimental situation develops, before we can
make some more definite statements.

Calculations of threshold corrections in realistic string theories are
very difficult and tedious. They were for along time only available for
very simple models, and turned out to be numerically very small[46].
This had lead to the  impression[49] that maybe a successful string
unification might necessarily require the introduction of new
particles at an intermediate scale ($10^{11}-10^{13}$GeV) that is much
smaller than the string scale and even the grand unified scale.

More recently, a breakthrough has been achieved in the calculation of the
moduli dependence of thresholds in (0,2) string theories[50].
When applied to realistic extensions of the supersymmetric standard model,
they show the possibility that the correct prediction of the low
energy coupling constant can be achieved without the introduction of a
small intermediate scale[51]. String unification is thus a
realistic possibility.

\bigskip\bigskip


\line{\bf 9. Conclusions\hfill}
\medskip
We have seen that the supersymmetric  model
provides an interesting framework for physics beyond the standard model.
In contrast to the standard model itself it might even have a simple
grand unified extension. Unfortunately up to now the model remains
a theoretical dream. No sign of supersymmetry has been detected yet.
We did not have the time here to discuss the experimental limits for
the various superpartners. Such a discussion can be found in [52]
with the yearly updates given in the big conferences. Of course,
still plenty of parameter space remains unexplored and we have to keep
in mind, that also the Higgs boson of the standard model has not been
found. So we have to  wait and see.

On the more theoretical side there could come some progress as well. I had no
time to discuss these developments in the lectures at this school and will
give an account of these issues elsewhere. Among the
much discussed subjects is the embedding of the supergravity models in the
framework of string theory. This might lead to more detailed information
on the nature and the size of the soft breaking terms, also in connection with
the mechanism of supersymmetry breakdown via gaugino condensates. Stringy
symmetries like so-called targed space duality could play an important role
in this process. For review and a list of references see ref. [53,54].

In these lectures, I concentrated on the simplest model with an exact
R-parity. This leads to a stable particle that might have cosmological
relevance. But there
are alternatives[55]. Of course, such models then necessarily
will have some amount of L(epton-number)-violation in dimension
four operators and it is not clear whether we would like to have those.
Alternative choices of discrete symmetries might avoid a possible
problem with the dimension-5 operators in grand unified models [38,39]
at the expense of L-violation. May be this could be relevant in connection
with the solar neutrino problem[56] as well as many
particle physics and cosmological phenomena.
\bigskip\bigskip


\line{\bf Acknowledgements\hfill}
\medskip
It is a pleasure to thank the organizers of the meeting for their
hospitality. This work was partially supported by  Deutsche
Forschungsgemeinschaft SFB-375-95 as well as grants from EC-contracts
SC1-CT91-0729 and SC1-CT92-0789.
\bigskip\bigskip

\line{\bf References\hfill}
\medskip

\item{1.}  S. Glashow, Nucl. Phys. B22 (1961) 579;
\item{ }   A. Salam, in Elementary Particle
     Theory, ed. N. Svartholm, (1968) 367;
\item{ }   S. Weinberg, Phys. Rev. Lett. 419 (1967) 1264

\item{2.}  M. Kobayashi and M. Maskawa,
     Progr. Theor. Phys. 49 (1973) 652

\item{3.}  E. Farhi and L. Susskind,
     Phys. Rep. 74C (1981) 277 and references therein

\item{4.}  J. Wess and J. Bagger, Supersymmetry, Princeton Series in Physics,
           Princeton Univ. Press (1983)

\item{5.}  H.P. Nilles,
     Phys. Rep. 110C (1984) 1

\item{6.}  P. Fayet, Nucl. Phys. B90 (1975) 104;
\item{ }  A. Salam and J. Strathdee,
     Nucl. Phys. B87 (1975) 85

\item{7.}  R.D. Peccei and H.R. Quinn,
     Phys. Rev. Lett. 38 (1977) 1440;
\item{ }  S. Weinberg, Phys. Rev. Lett. 40 (1978) 223;
\item{ }  F. Wilczek, Phys. Rev. Lett. 40 (1978) 229

\item{8.}  J.E. Kim,
     Phys. Rev. Lett. 43 (1979) 103

\item{9.}  D.Z. Freedman, S. Ferrara and P. van Nieuwenhuizen,
     Phys. Rev. D13 (1976) 3214

\item{10.} For a review see: P. van Nieuwenhuizen,
     Phys. Rep. 68C (1981) 189

\item{11.}  E. Cremmer, S. Ferrara, L. Girardello and A. van Proeyen,
     Nucl. Phys. B212 (1983) 413

\item{12.} For a review see: H.P. Nilles,
     International Journal of Modern Physics A5 (1990) 4199

\item{13.} S. Ferrara, L. Girardello and H.P. Nilles,
     Phys. Lett. 125B (1983) 457

\item{14.}  E. Witten,
     Phys. Lett. 155B (1985) 151

\item{15.} J. Ellis, A.B. Lahanas, D.V. Nanopoulos and K. Tamvakis,
     Phys. Lett. 134B (1984) 429

\item{16.}  H.P. Nilles, M. Srednicki and D. Wyler,
     Phys. Lett. 120B (1983) 346

\item{17.}    H.P. Nilles, Phys. Lett. 115B (1982) 193;
              Nucl. Phys. B217 (1983) 366

\item{18.}  J.P. Derendinger, L.E. Ibanez and H.P. Nilles,
     Phys. Lett. 155B (1985) 65;
\item{ }     M. Dine, R. Rohm, N. Seiberg and E. Witten,
     Phys. Lett. 156B (1985) 55

\item{19.} J.E. Kim and H.P. Nilles,
     Phys. Lett. 138B (1984) 150

\item{20.} K. Inoue, A. Kakuto, H. Komatsu and S. Takeshita,
     Progr. Theor. Phys. 68 (1982) 927

\item{21.}  H. P. Nilles, Tasi lectures 1990, Published in the proceedings,
            Ed. P. Langacker, page 633

\item{22.}  P. Langacker, Phys. Rep. 72C (1981) 185

\item{23.}  S. Dimopoulos, S. Raby and F. Wilczek,
     Phys. Rev. D24 (1981) 1681

\item{24.} P. Langacker and M.-X. Luo, Phys. Rev. D44 (1991) 817;
\item{   } J. Ellis, S. Kelley and D. Nanopoulos, Phys. Lett. B249 (1990) 441;
\item{   } U. Amaldi, W. De Boer and H. F\"urstenau, Phys. Lett. B260 (1991)
447

\item{25.}  E. Witten, Phys. Lett. 105B (1981) 267;
  \item{ }   L.E. Ibanez and G.G.Ross, Phys. Lett. 110B (1982) 215

\item{26.}  H.P. Nilles, M. Srednicki and D. Wyler,
     Phys. Lett. 124B (1983) 337

\item{27.} B. Grinstein,
     Nucl. Phys. B206 (1982) 387

\item{28.}  S. Weinberg, Phys. Rev. D26 (1982) 287;
\item{ }     N. Sakai and T. Yanagida, Nucl. Phys. B197 (1982) 533

\item{29.}  P. Langacker and N. Polonsky, Phys. Rev. D47 (1993) 4028

\item{30.}  M. Carena, S. Pokorski and C. Wagner, Nucl. Phys. B406 (1993) 59

\item{31.}  V. Barger, M. S. Berger and P. Ohmann, Phys. Rev. D47 (1993) 1093

\item{32.}   A.J. Buras, J. Ellis, M.K. Gaillard and D.V. Nanopoulos,
     Nucl. Phys. B135 (1978) 66

\item{33.}   L.E. Ibanez and C. Lopez, Nucl. Phys. B233 (1984) 511

\item{34.}  H. Arason, D.J. Castano, B. Keszthelyi, S. Mikaelian,
        E. J. Piard, P. Ramond and B.D. Wright, Phys. Rev. Lett. 67 (1991) 2933

\item{35.}  C. T. Hill, Phys. Rev. D24 (1981) 691

\item{36.}  R. Arnowitt and P. Nath, Phys. Rev. Lett. 69 (1992) 725;
            Phys. Lett. B287 (1992) 89

\item{37.}  J. Hisano, H. Murayama and T. Yanagida, Nucl. Phys. B402 (1993) 46

\item{38.} L. E. Ibanez and G. G. Ross, Phys. Lett. B260 (1991) 291;
           Nucl. Phys. B368 (1992) 3

\item{39.} P. Mayr, D. Kapetanakis and H. P. Nilles, Phys. Lett. B282 (1992) 95

\item{40.} See  M. Olechowski and S. Pokorski, Nucl. Phys. B404 (1993) 590
           for a discussion and further references

\item{41.} K. Hagiwara and Y. Yamada, Phys. Rev. Lett. 70 (1993) 709

\item{42.} D. Matalliotakis and H. P. Nilles, Nucl. Phys. B435 (1995) 115

\item{43.}  L. E. Ibanez and H. P. Nilles, Phys. Lett. 169B (1986) 354

\item{44.}  V. Kaplunovsky, Nucl. Phys. B307 (1988) 145 and Stanford
            preprint ITP-838, revised version (May 1992)

\item{45.}  L. Dixon, V. Kaplunovsky  and J. Louis, Nucl. Phys. B355
            (1991) 649

\item{46.}  P. Mayr and S. Stieberger, Nucl. Phys. B407 (1993) 725

\item{47.}  P. Mayr, H.P. Nilles and S. Stieberger, Physics Letters B317
            (1993) 53

\item{48.}  L. E. Ibanez, D. L\"ust and G. G. Ross, Phys. Lett. B272
            (1991) 251
\item{   }  L. E. Ibanez and D. L\"ust, Nucl. Phys. B382 (1992) 305

\item{49.}  K. Dienes and A. Faraggi, preprint IASSNS-HEP-95-24,
            hep-th/9505018

\item{50.}  P. Mayr and S. Stieberger, Phys. Lett. B355 (1995) 107

\item{51.}  H. P. Nilles and S. Stieberger, preprint TUM-HEP-225-95,
            hep-th/9510009

\item{52.} Phenomenological aspects of supersymmetry, Lecture notes in
           Physics 405, Eds. W. Hollik, R. R\"uckl and J. Wess,
           Springer-Verlag Berlin Heidelberg 1992

\item{53.}  S. Ferrara and S. Theisen, CERN TH-5652 (1990), Lectures at
            the third Hellenic Summer School, Corfu, Greece, published
            in the Proceedings, page 620

\item{54.} M. Cvetic, A. Font, L.E. Ibanez, D. L\"ust and
           F. Quevedo, Nucl. Phys. B361 (1991) 194

\item{55.}  J.W.F Valle, Prog. Part. Nucl. Phys. 26 (1991) 91 and
            references therein

\item{56.}  E. Roulet, Phys. Rev. D44 (1991) 935;
\item{   }  M. M. Guzzo, A. Masiero and S. T. Petcov. Phys. Lett. B260
            (1991) 154

\end